\documentstyle[aps,psfig,multicol]{revtex}
\newcommand{\ignore}[1]{}
\begin{document}
\def\be{\begin{equation}}
\def\ee{\end{equation}}
\def\ba{\begin{eqnarray}}
\def\ea{\end{eqnarray}}

\title{Odd-even binding effect from random two-body interactions}
\author{Thomas Papenbrock$^1$, Lev Kaplan$^2$, and George F. Bertsch$^2$}
\address{$^1$Physics Division, Oak Ridge National Laboratory, Oak Ridge,
 Tennessee 37831}
\address{$^2$Institute for Nuclear Theory,
 University of Washington, Seattle, Washington 98195}
\maketitle
\begin{abstract}
Systematic odd-even binding energy differences in finite metallic particles are
usually attributed to mean-field orbital energy effects or to a coherent
pairing interaction.  We show analytically and numerically that a purely random
two-body Hamiltonian can also give rise to an odd-even staggering.  We
explore the characteristics of this chaotic mechanism and discuss
distinguishing features with respect to the other causes of staggering. In
particular, randomness-induced staggering is found to be a smooth function of
particle number, and the mechanism is seen to be largely insensitive to the
presence of a magnetic field.
\end{abstract}

\pacs{PACS numbers: }

\begin{multicols}{2}

\section{Introduction}
Interacting finite fermionic systems such as atomic nuclei, metallic
clusters\cite{de93}, and small metallic grains\cite{BRT} display an odd-even
staggering in ground-state energies, i.e., the binding energy of an even-number
system is larger than the arithmetic mean of its odd-number neighbors. There
are two well-known mechanisms that can give rise to this staggering, namely the
Kramers degeneracy in the mean-field Hamiltonian and the BCS mechanism arising
from an attractive effective interaction. In nuclei, the BCS pairing mechanism
resulting from a residual nucleon-nucleon interaction is dominant\cite{BM}, but
the mean-field or orbital energy effect may also be significant in the lighter
nuclei\cite{satula}. Surprisingly, many basic phenemona normally associated
with pairing can also arise from random interactions. The behavior of
random-interaction ensembles has mostly been studied in a nuclear physics
context \cite{JBD,Kusnezov,Zel,Drozdz,Arima,BF,jo99} but there has also
been some study of spectra in the context of small metallic grains\cite{Jac}.  

In the case of metallic clusters of fewer than a hundred atoms, the orbital
energy effect is rather strong and staggering is seen for species that do not
exhibit superconductivity.  This effect can be easily understood using a
jellium model or density functional theory\cite{ma94}.  On the other hand, the
staggering effect seen in Ref.~\cite{BRT} may have some contribution from the
BCS pairing mechanism.  A number of theoretical studies have been
made~\cite{pairmod} using techniques applicable to large finite
systems~\cite{Rich}.  Taking a uniform mean-field spectrum and an attractive
pairing interaction with constant coupling, one observes a smooth crossover
from BCS superconductivity in the bulk to the few-electron regime.  For small
systems, the gap is of the size of the mean level spacing and thus ceases to be
an indicator for pairing.  Nevertheless, strong pairing correlations and
odd-even staggering persist as the system size decreases.

In a grain with irregular boundaries, one expects that the electron
orbitals will have a chaotic character and therefore the interaction
will have a random as well as a regular part.  In this paper we will
introduce such an interaction and study its typical effects on the
binding systematics.  Our Hamiltonian thus includes attractive and
repulsive pairing interactions as well as more general two-body
interactions. The assumption of randomness is motivated as follows:
For nuclei it is well known that the residual interaction leads to
fluctuation properties in wave functions and energy levels that are
similar to those of random matrices taken from the Gaussian orthogonal
ensemble \cite{ZBFH}. In the case of small metallic grains or quantum
dots, one may assume that their irregular shape leads to chaoticity in
the single-particle wave functions \cite{Alhassid}.  This in turn
causes randomness in those two-body matrix elements that link four
different orbitals with each other. Matrix elements between pairs of
orbitals that are related by time-reversal symmetry need not
necessarily be random, and these determine the ``coherent'' terms of
the interaction.

A realistic Hamiltonian for quantum dots or small metallic grains would thus
conserve total spin and include spin-independent one-body terms, random
two-body interactions, and coherent interactions that are non-random but have
attractive and repulsive components.  The most general Hamiltonian to study
generic properties when all these features are included may be written as
\begin{eqnarray}
H &=& \sum_{i,\sigma} \varepsilon_i \, c^\dagger_{i\sigma}c_{i\sigma} 
+\sum_i (u+ u'_{i}) \, c^\dagger_{i\uparrow}c^\dagger_{i\downarrow}
c_{i\downarrow}c_{i\uparrow} \nonumber \\&+&
\sum_{ij} \sum_{\sigma_1\sigma_2\sigma_3\sigma_4}
[(w_0 +w'_{0,ij})\langle \sigma_1 |\sigma_2\rangle
\langle \sigma_3 |\sigma_4\rangle  \nonumber \\
&&+(w_1 +w'_{1,ij})\langle \sigma_1 |\vec\sigma|\sigma_2\rangle\cdot
\langle \sigma_3 |\vec\sigma|\sigma_4\rangle] \, 
c^\dagger_{i\sigma_1}c^\dagger_{j\sigma_2}
c_{j\sigma_3}c_{i\sigma_4} \nonumber \\ &+&
\sum_{ij} (g +g'_{ij}) \,
c^\dagger_{i\uparrow}c^\dagger_{i\downarrow}
c_{j\downarrow}c_{j\uparrow} \nonumber \\ &+&
\sum_{ijkl} \sum_{\sigma_1\sigma_2\sigma_3\sigma_4}
[v_{0,ijkl}\langle \sigma_1 |\sigma_2\rangle
\langle \sigma_3 |\sigma_4\rangle  \nonumber \\
& & +v_{1,ijkl}\langle \sigma_1 |\vec\sigma|\sigma_2\rangle\cdot
\langle \sigma_3 |\vec\sigma|\sigma_4\rangle] \,
c^\dagger_{i\sigma_1}c^\dagger_{j\sigma_2}
c_{k\sigma_3}c_{l\sigma_4} \,.
\end{eqnarray}

Here the coherent parts of the interaction are represented by the
terms with coefficients $u$, $w_s$, and $g$.  The fluctuating parts of
the interaction are represented by the terms containing $u'$, $w'_s$,
$g'$, and $v_s$. These fluctuating parts are typically taken from
ensembles with a Gaussian distribution; they are thus characterized by
the width of the Gaussian.  The single-particle term $\varepsilon_i$
sets the energy scale and may often be taken to give a uniform spacing
of levels without loss of generality.  This full Hamiltonian is
difficult to study due to its many parameters. There have been many
studies in the limit in which fluctuation effects are only included in
the single-particle Hamiltonian
$\varepsilon_i$\cite{ullmo,sh01,be01,sm96}. We consider a very
different limit, neglecting the coherent terms in the interaction and
assuming the $v_s$ term to dominate the fluctuating parts.  Properties
of such random two-body interaction ensembles have been studied
extensively in nuclear physics\cite{bo70,fr70,be01b,Kota}.

When the Hamiltonian of the nuclear shell model was modeled in this way, it was
found that the spectral properties were quite regular for the ground states.
As examples we mention $J^P=0^+$ ground-state dominance in shell model
calculations with random interactions \cite{JBD,Kusnezov,Zel,Drozdz,Arima},
band structure in interacting boson models with random couplings \cite{BF},
structure in ground-state wave functions of two-body random ensembles
\cite{KP}, and an odd-even binding effect in filling a large shell\cite{jo99}.
In the context of quantum dots, the random two-body interactions were found 
strongly to favor singlet ground-state spins \cite{Jac,KPJ}. Recently, this
structure has been investigated using the group symmetry of the random
Hamiltonians ~\cite{KK}. These findings suggest that the structure of
interacting many-body systems is to some extent already determined by the rank
of the interaction alone, and one does not need all the details of the
interaction. We will show that odd-even staggering also fits into this picture
and is not solely a consequence of an attractive pairing force. 

This paper is organized as follows. In Section~\ref{mf} we introduce the
Hamiltonian and discuss the odd-even effects arising from the one-body part
alone. Section~\ref{2body} contains analytical results for the odd-even
staggering due to a random two-body interaction (some technical details of this
analytical analysis are included in an Appendix).  The crossover between the
mean-field regime and the regime of strong interactions is numerically
investigated in Section~\ref{trans}. The effects of breaking time-reversal
symmetry are studied in Section~\ref{time}. Finally, we give a summary.  

\section{Hamiltonian and staggering indicator}
\label{mf}

As discussed in the introduction, we will consider ensembles of Hamiltonians
including only a single-particle energy and a random two-body interaction.  We
write this in the form
\begin{eqnarray}
H&=&\sum_{i=1}^M \varepsilon_i\,(c^\dagger_{i\uparrow}c_{i \uparrow}+
c^\dagger_{i\downarrow}c_{i \downarrow}) \nonumber \\ &+&
C_0 \sum_{\alpha,\alpha' \; {\rm spin}-0\; {\rm pairs}}
v_{0\alpha\alpha'} A^\dagger_\alpha
A_{\alpha'} \nonumber \\
&+&C_1 \sum_{\beta,\beta' \; {\rm spin}-1\; {\rm pairs}}
 v_{1\beta\beta'} A^\dagger_\beta
A_{\beta'}\,.
\label{hamilt}
\end{eqnarray}
The first term represents the mean-field contribution, where $\varepsilon_i$ is
the single-particle energy associated with orbital $i$, and $c_{i\uparrow}$,
$c_{i\downarrow}$ are the one-particle annihilation operators for that orbital.
As usual, we assume an ordering $\varepsilon_i \le \varepsilon_{i+1}$. The
second and third terms represent the interaction for pairs having spin $S$
equal to 0 and 1, respectively.  The operators $A_\alpha$ in the second term
are spin-singlet two-particle annihilation operators $A_\alpha =
(c_{i\downarrow}c_{j\uparrow} - c_{i\uparrow}c_{j\downarrow})/
\sqrt{2(1+\delta_{ij})}$ with $\alpha$ standing for the set of orbital pairs
$ij$.  The $A_\beta$ in the third line are similarly defined for spin-triplet
pairs.

The randomness assumption tells us that there is no preferred basis within
either the $S=0$ or $S=1$ sector of two-body states. The couplings $v_{s\alpha
\alpha'}$ then should be taken from the Gaussian orthogonal random-matrix
ensemble (GOE).  We fix the variance of the $v_s$ to be unity for off-diagonal
elements.  The GOE then satisfies
\begin{equation}
\label{goe1}
\langle v^2_{0\alpha \alpha'} \rangle  = 1 + \delta_{\alpha \alpha'}\,,
\end{equation}
where $\langle \cdots \rangle$ indicates an ensemble average and similarly for
$v_{1\beta\beta'}$.  We are concentrating for now on the case of time-reversal
symmetry, so the matrices $v_0$ and $v_1$ are real and symmetric. The case of
broken time-reversal symmetry in the presence of a magnetic field will be 
considered in Section~\ref{time}.

The prefactors $C_0$ and $C_1$ allow us to consider arbitrary strengths of the
spin-0 and spin-1 couplings relative to each other and relative to the
single-particle level spacing. As we will see below, several qualitatively
different regimes for ground-state staggering are possible within this simple
random model, depending on the values $C_0$ and $C_1$ as well as on particle
density.

Let us denote the ground-state energy of the $N$-body system as $E(N)$.  A
useful staggering indicator is the empirical pairing gap
\begin{equation}
\label{gap}
\Delta(N) \equiv {1\over 2}[E(N+1)-2 E(N) +E(N-1)]\,.
\end{equation}
This three-point observable is essentially the ``curvature" or second
derivative of the binding energy with respect to particle number $N$.  Positive
(negative) $\Delta(N)$ indicates that the binding energy of the $N$-body system
is larger (smaller) than the arithmetic mean of the binding energies of its
neighbors. We have an odd-even staggering whenever $\Delta(N)$ staggers with
$N$.

It is instructive to consider the trivial case where residual interactions are
negligible, i.e., $C_0=C_1=0$. Then the $N-$particle ground-state energy is
given by $E(N)=2\sum_{i=1}^{N/2} \varepsilon_i$ for $N$ even and
$E(N)=E(N-1)+\varepsilon_{(N+1)/2}$ for $N$ odd. Here $N$ may range between $0$
and $2M$, where $M$ is the number of available orbitals.  One obtains for the
empirical pairing gap 
\ba
\label{meanfield}
\Delta(N)=\left\{
     \begin{array}{ll} 
     (\varepsilon_{(N/2)+1}-\varepsilon_{N/2})/2 \ge 0& \mbox{for $N$ even,}\\
      0 & \mbox{for $N$ odd.}
     \end{array}
     \right.
\ea
Thus, there is a trivial odd-even staggering due to the mean-field alone.  In
what follows we will mainly be interested in the effects of interactions, and
in the effects of adding a magnetic field. For odd-number systems, a nonzero
value of the empirical pairing gap must be due to interactions, and this allows
one easily to discriminate mean-field effects from interactions. Such a
discrimination is more difficult for even-number systems and has recently been
studied in mean-field plus pairing Hamiltonians \cite{satula,Doba,Heenen}. We
will see in Section~\ref{trans} how mean-field effects can be distinguished
from staggering caused by complex (or random) interactions.  Note that an
electric charging energy $E_{\rm charge} = c N(N-1)$ leads only to a
$N$-independent constant shift $\Delta(N)\rightarrow \Delta(N)+c$ and can
therefore be neglected.

\section{Effects from random two-body interactions}
\label{2body}
We now imagine the opposite situation from that of the previous section, i.e.,
we consider the regime $\varepsilon_i=0$ where mean-field effects are
negligibly weak compared with the random two-body interaction. In this limit
one might assume that all odd-even effects should disappear. Surprisingly, this
turns out not to be the case. Instead, we find persistent odd-even staggering
arising only from the random two-body interactions; stronger binding energies
for even-$N$ systems are typically obtained in numerical simulations.

To understand this result analytically, we first note that the spectral density
of a system with two-body interactions approaches a Gaussian shape in the 
many-body limit $N \to \infty$ \cite{Gervots,Mon}. The ground-state energies
for different particle number or spin sectors are then largely determined by
the widths $\sqrt{{\rm Tr} H^2}$ of the corresponding Gaussians, scaling as 
\begin{equation}
E \approx b\sqrt{{\rm Tr} H^2} \,,
\label{proport}
\end{equation}
where it is assumed without loss of generality that ${\rm Tr} H=0$.  The
prefactor $b$ depends of course on the details of the deviations of the
spectral shape from an exact Gaussian form, since these deviations cut off the
tails of the Gaussian. Following an analysis along the lines of
Ref.~\cite{Gervots}, where the spectral shape is expanded in terms of Hermite
polynomials, and then estimating the coefficients of these polynomials, one may
conjecture that the prefactor $b$ should scale as $\log N$ with the number of
particles in the system. In any case, for our purposes it is sufficient that
this prefactor varies smoothly with $N$ without significant staggering, which
is confirmed by numerical simulations. Eq.~(\ref{proport}) is known to provide
a good qualitative explanation for some observed behavior of low-lying spectra,
even for moderate numbers of particles where the Gaussian approximation is far
from valid. For example, a comparison of ${\rm Tr} H^2$ for different spin
sectors helps to explain $J=0$ total spin dominance among the ground states of
random interacting many-body systems~\cite{Jac,KPJ}.

\subsection{Dilute limit}

Applying this approach to the present problem, we need then to understand
how ${\rm Tr}H^2$ depends on the number of particles and other parameters
of the system. For simplicity, we consider first the dilute limit $N \ll M$
with a pure $S=0$ two-body coupling ($C_1=0$).

From previous work, it is known that for even $N$ the ground state comes always
from the sector of total spin $J=0$. In the dilute limit, a typical basis state
in this sector has the form
\begin{equation}
|\Psi_{J=0}\rangle = 2^{-N/2} \prod_{z=1}^{N/2}
(a^\dagger_{i_z\downarrow}a^\dagger_{j_z \uparrow}-
a^\dagger_{i_z\uparrow}a^\dagger_{j_z \downarrow})|0\rangle\,,
\end{equation}
where the $N$ orbitals $i_z$, $j_z$ are all distinct. One easily checks that
the number of $S=0$ pairs in this state is $(N^2+2N)/8$, since the particles on
orbitals $i_z$ and $j_z$ for a given $z$ are in an $S=0$ combination by
construction, while the remaining  $(N^2-2N)/2$ pairs have a probability $1/4$
of being in a singlet combination. Any of these $S=0$ pairs, labeled by
$\alpha'$ in Eq.~(\ref{hamilt}), may be annihilated by the $C_0$ term in the
Hamiltonian.  Another $S=0$ pair, $\alpha$, must then be created; there are
$M^2/2$ choices for $\alpha$ in the dilute limit. Thus, simply by counting the
number of terms in the $C_0$ part of the Hamiltonian in Eq.~(\ref{hamilt}) that
may act on a total spin $J=0$ basis state we find
\begin{equation}
{\rm Tr}_{( {\rm even}\; N \ll M)} H^2 = C_0^2 {M^2 (N^2+2N) \over 8}
\end{equation}
for $N$ even and $N \ll M$.

Similarly, for odd $N$ the preferred many-body ground state has total spin
$J=1/2$. The typical basis state has the form
\begin{equation}
|\Psi_{J=1/2}\rangle = 2^{-N/2} a^\dagger_{k\uparrow} \prod_{z=1}^{N/2}
(a^\dagger_{i_z\downarrow}a^\dagger_{j_z \uparrow}-
a^\dagger_{i_z\uparrow}a^\dagger_{j_z \downarrow})|0\rangle\,,
\end{equation}
where we take $J_z=+1/2$ without loss of generality, and the indices $i_z$,
$j_z$, and $k$ are all distinct. This state contains only $(N^2+2N-3)$ singlet
pairs, resulting in
\begin{equation}
{\rm Tr}_{({\rm odd} \; N \ll M)} H^2 = C_0^2 {M^2 (N^2+2N-3) \over 8}\,.
\end{equation}
The $O(1/N^2)$ difference in the widths explains the odd-even staggering in
ground-state energies. Intuitively, the result is easy to understand: the
ground state of the odd-$N$ system is forced to have a slightly higher total
spin, resulting in a slightly smaller fraction of spin-$0$ pairs and
consequently a smaller effect of the $C_0$ term in the Hamiltonian. This in the
end is what leads to weaker binding for the odd-$N$ system.

The above analysis also gives a quantitative prediction for the size of the
staggering effect. Assuming in accordance with Eq.~(\ref{proport}) that the
ratio of ground-state energies is proportional to the ratio of the widths, we
find
\begin{equation}
|E_{{\rm even} N}|=|E_{odd N}|\left(1+{3 \over 2N^2}\right) 
\end{equation}
for large $N$ in the dilute limit and therefore
\begin{equation}
\Delta(N)_{C_0}=(-1)^N {3 \over 2 N^2}|E(N)| 
\label{dilutedelta}
\end{equation}
to leading order.  We may compare this with the size of the pairing gap for the
mean-field dominated system. In the previous section, we saw that $\Delta(N)
=\Delta/2$ on average for $N$ even, where $\Delta$ is the mean level spacing of
the single-particle spectrum. This can be normalized, however, in units of the
binding energy.  This binding energy, i.e., half the many-body spectral width,
is $|E| \approx MN \Delta/2$ in the mean-field case. So the average pairing gap
has the size
\begin{equation}
\Delta(N)_{{\rm mean-field}}={1 \over MN}|E(N)|
\end{equation}
for even $N$, which surprisingly is {\it smaller} than the pure
interaction-induced pairing gap in the dilute limit $N \ll M$.

At finite particle density $\rho=2N/M$, mean-field-induced and
interaction-induced stagger are of comparable size, a characteristic difference
being the vanishing of the pairing gap $\Delta(N)$ for odd $N$ in the
mean-field case, Eq.~(\ref{meanfield}), which is absent for the pure
interacting theory.  In addition, in the presence of fluctuations in the
single-particle spectrum, mean-field induced $\Delta(N)$ will itself fluctuate
between successive even values of $N$, while interaction-induced stagger is
predicted to be smooth.  These analytic predictions will be verified
numerically in Sec.~\ref{trans} below.

\subsection{General results for finite density}

The above derivation, though strictly valid only in the dilute limit, in fact
provides a correct intuitive explanation of the stagger at any density for a
pure $S=0$ two-body interaction. Handling the $S=1$ interaction requires more
care, since the qualitative behavior will depend strongly on the density
$\rho$.  We therefore need the exact expressions for ${\rm Tr} H^2$ in various
particle number and spin sectors. These expressions may be straightforwardly,
though perhaps rather tediously, obtained by applying the original Hamiltonian,
Eq.~(\ref{hamilt}), to various basis states and evaluating the norm.

The full results are presented in the Appendix. There we find that for a pure
singlet random interaction, the prediction of Eq.~(\ref{dilutedelta}) for the
size of the staggering, obtained above only in the dilute limit, is in fact
confirmed as a lower bound for arbitrary densities in the many-body limit
$N \to \infty$:
\begin{equation}
(-1)^N \Delta(N)_{C_0} \ge {3 \over 2 N^2}|E(N)| \,.
\end{equation}

The situation is more complex for a pure triplet coupling ($C_0=0$), since here
the ground state may be a state of either minimal or maximal spin. In this case
we see using formulas given explicitly in the Appendix that a critical
density $\rho_{crit}$ exists below which there is no staggering, while above
which interaction-induced staggering of order $|E(N)|/N^2$ appears, just
as in the singlet case. As the singlet coupling is turned on, $\rho_{crit}$
decreases, reaching $0$ at $C_0=C_1$. Thus, odd-even staggering with stronger
binding for even-$N$ systems is predicted to be a very general consequence of
random two-body interactions, present for pure-singlet and pure-triplet
interactions as well as in the intermediate case.

\section{Crossover between Mean-Field regime and strong Two-Body Interactions}
\label{trans}

The analytical results of the previous sections were obtained for pure
one-body or pure two-body interactions. In this section we will study the
odd-even staggering for the full Hamiltonian~(\ref{hamilt}) numerically.  To
this purpose we draw the random matrices $v_{0}$ and $v_{1}$ in
Eq.~(\ref{hamilt}) from the GOE and compute the ground-state energies of
Hamiltonian of Eq.~(\ref{hamilt}) for several particle numbers $N$. This
procedure is repeated many times for each $N$ to obtain ensemble-averaged
values for the ground state energies $E(N)$ and the empirical pairing gap
defined in Eq.~(\ref{gap}).  In what follows we set the number of
single-particle orbitals to $M=10$, and obtain ensemble averages from 200 runs.
The largest matrices of the ensemble have dimension $63504$; their ground
states are computed using the sparse matrix solver {\sc Arpack} \cite{arpack}.

\begin{figure}[h]
  \begin{center}
    \leavevmode
    \parbox{0.5\textwidth}
           {\psfig{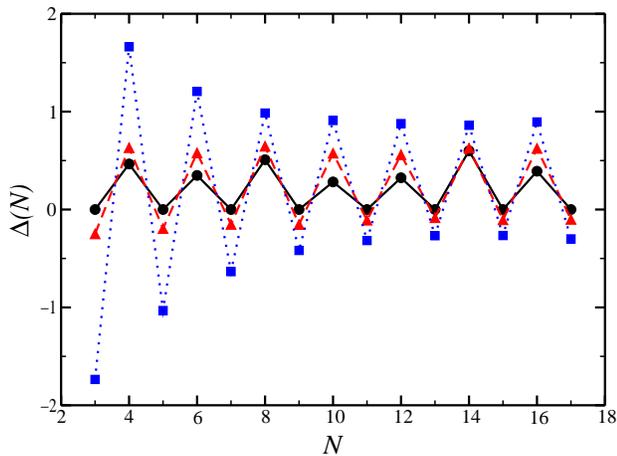}}
  \end{center}
\protect\caption{Empirical pairing gap as a function of particle number for
parameter $\varphi=0$ (full line), $\varphi=\pi/12$ (dashed line), and
$\varphi=\pi/2$ (dotted line; graph scaled by a factor $1/2$ for display
purposes) shows the transition from the mean-field regime to strong
interactions in the spin-0 channel.}
\label{fig1}
\end{figure}

We have to assign values to the single-particle energies $\varepsilon_i$ of the
mean field and to the coupling constants $C_0$ and $C_1$ of the two-body
interactions. We assume a mean-field spectrum with level spacings
$\varepsilon_{i+1}-\varepsilon_i$ that are  Wigner-distributed. This is
consistent with the assumption that our quantum dot or metallic grain has
irregular shape. To study the transition, we multiply the single-particle
energies with a factor $\cos{\varphi}$ and set the spin-0 coupling
$C_0(\varphi)=\sin{\varphi}$. Here $\varphi$ is in the range $\varphi\in
[0,\pi/2]$ and thus parameterizes the transition from the mean field to the
regime of strong interactions.  The spin-1 coupling $C_1$ is set to zero.
Figure~\ref{fig1} shows the empirical pairing gap~(\ref{gap}) as a function of
particle number $N$ for parameter values $\phi=0, \pi/12, \pi/2$. 

We see from Figure~\ref{fig1} that the odd-even staggering persists throughout
this transition. In the absence of the mean field ($\phi=\pi/2$), the
staggering  decreases slowly with increasing $N$ and then increases again very
close to the maximal filling, when the number of holes becomes small and $\rho$
approaches unity in Eq.~(\ref{c0diff}). Its envelope depends smoothly on $N$ if
only even  or only odd values of $N$ are considered.  These qualitative results
are fully consistent with the analytical predictions obtained in
Sec.~\ref{2body} and in the Appendix.  The absence of such a smooth envelope
thus indicates that the staggering is instead dominated by mean-field effects,
as in the $\varphi=0$ line in Figure~\ref{fig1}. Similar observations have been
made for pairing-plus-quadrupole in Ref.\cite{Doba}.  Note that the random
interactions drive the empirical pairing gap $\Delta(N)$ to negative values for
odd $N$; in this sense the staggering is more pronounced in the presence of
interactions than in the mean-field regime. Note also that the magnitude of the
staggering itself contains only little information since the transition from
the noninteracting to the interacting Hamiltonian does not correspond to a
transition in a physical system. 
 
\begin{figure}[h]
  \begin{center}
    \leavevmode
    \parbox{0.5\textwidth}
           {\psfig{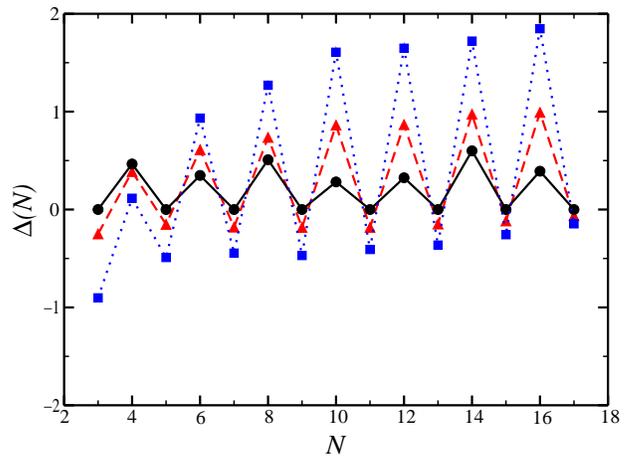}}
  \end{center}
\protect\caption{Empirical pairing gap as a function of particle number
for parameter $\varphi=0$ (full line), $\varphi=\pi/12$ (dashed line), and
$\varphi=\pi/2$ (dotted line; graph scaled by a factor $1/2$ for display
purposes) shows the transition from the mean-field regime to strong
interactions in the spin-1 channel.}
\label{fig2}
\end{figure}

We repeat these calculations in Fig.~\ref{fig2} for the case of vanishing
spin-0 coupling, $C_0=0$, and set the spin-1 coupling to
$C_1(\varphi)=\sin{\varphi}$.  Again, odd-even staggering persists throughout
the transition. In the regime of strong interactions the magnitude of the
empirical pairing gap increases with increasing $N$ for even $N$.  The
situation is reversed for odd values of $N$. Leaving out very small systems
($N=3$), the envelopes for even and odd $N$ are still smooth enough to
discriminate mean-field effects from interaction-induced pairing.

\begin{figure}[h]
  \begin{center}
    \leavevmode
    \parbox{0.5\textwidth}
           {\psfig{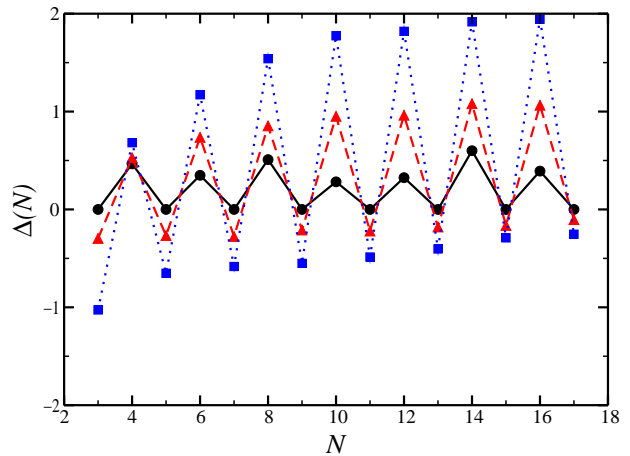}}
  \end{center}
\protect\caption{Empirical pairing gap as a function of particle number for
parameter $\varphi=0$ (full line), $\varphi=\pi/12$ (dashed line), and
$\varphi=\pi/2$ (dotted line; graph scaled by a factor $1/2$ for display
purposes) shows the transition from the mean-field regime to strong
interactions.}
\label{fig3}
\end{figure}

Finally, we consider the case of equally strong spin-0 and spin-1 couplings and
set $C_0(\varphi)=C_1(\varphi)=\sin{\varphi}$.  Figure~\ref{fig3} shows that
this case is qualitatively similar to the case of pure spin-1 coupling, since
triplet pairs outnumber singlet pairs by a 3:1 ratio in the large-$N$ limit.
Again, the interaction-induced staggering exhibits a smooth envelope and can
therefore clearly be distinguished from mean-field effects.

\section{Magnetic Field Effects}
\label{time}

BCS-like pairing results from strong correlations between fermions in
time-reversed orbitals. Thus, these correlations can be destroyed by a
sufficiently strong breaking of time-reversal symmetry. Examples of
this well-known phenomenon are the breakdown of electronic
superconductivity in the presence of sufficiently strong magnetic
fields and the reduction of pairing correlations in rapidly rotating
and deformed nuclei. In this section we want to study how breaking
time-reversal symmetry affects the odd-even staggering in systems with
a random two-body interaction. Having metallic grains in mind we thus
consider the effect of a magnetic field. To be definite, we take a
uniform $B$-field in the $z$-direction. This leads to Zeeman splitting
and adds the following one-body term to the Hamiltonian
\be 
\label{Z}
H_B = \mu B\sum_{i=1}^M\left(c^\dagger_{i\uparrow}c_{i \uparrow}-
c^\dagger_{i\downarrow}c_{i \downarrow}\right), 
\ee 
which also breaks rotational symmetry, i.e., only the projection of
the total spin $J_z$ remains conserved. Here, $\mu$ is an appropriate
constant.  A second effect consists of the modification of the random
two-body interaction.  Provided the time-reversal symmetry breaking
induces splittings that are larger than the mean level spacing, the
random matrices $v_{0\alpha\alpha'}$ and $v_{1\beta\beta'}$ in the
Hamiltonian~(\ref{hamilt}) have to be drawn from the Gaussian unitary
ensemble (GUE). Accordingly, Eq.~(\ref{goe1}) for the $S=0$ matrix
$v_{0\alpha\alpha'}$ and the corresponding formula for the $S=1$
matrix $v_{1\beta\beta'}$ have to be replaced by
\be
\label{gue}
\langle |     v_{0\alpha \alpha'}|^2 \rangle = 
\langle | v_{1\beta \beta'}|^2 \rangle  = 1.
\ee
This reduces the variance of the diagonal matrix elements by a factor of two
when compared to the GOE. Considering the random two-body interaction alone,
this effect introduces only small corrections of order $1/N^2$ to the results
presented in the previous sections and in the Appendix. 

Let us consider the trivial case where residual interactions can be neglected.
The $B$-dependent pairing gap then becomes
\ba
\label{B}
\Delta(N,B)=\left\{ \begin{array}{ll}
     {1\over 2}\left(\varepsilon_{N/2+1}-\varepsilon_{N/2}\right) 
                     -\mu B &\mbox{for $N$
     even,}\\ 
     \mu B & \mbox{for $N$ odd.}  
     \end{array} 
     \right.  
\ea
The odd-even staggering thus decreases with increasing magnetic field and
disappears when the Zeeman splitting $2\mu B$ equals half the mean level
spacing $\langle\varepsilon_{i+1} - \varepsilon_{i}\rangle$. Note that
Eq.~(\ref{B}) ceases to be applicable for stronger magnetic fields. In the
limit of very large $B$-fields, the ground state becomes spin polarized (i.e.,
has maximal spin $J=N/2$) and any odd-even staggering disappears. Note also
that a breaking of time-reversal symmetry leads to a positive pairing gap at
odd $N$ and can thereby easily be distinguished from the effects of
interactions.

\begin{figure}[h]
  \begin{center}
    \leavevmode
    \parbox{0.5\textwidth}
           {\psfig{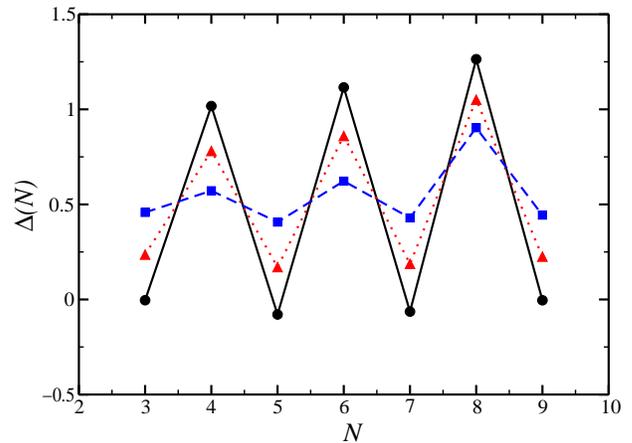}}
  \end{center}
\protect\caption{Empirical pairing gap as a function of particle number
for various strengths of the magnetic field: 
$2\mu B/\langle\varepsilon_{i+1}-\varepsilon_i\rangle = 0, 1/4, 1/2$
(full line, dotted line, dashed line).}
\label{fig4}
\end{figure}

We now include again the random two-body interactions and compute the empirical
pairing gap as the magnetic field is switched on. The number of single-particle
orbitals is $M=6$. At vanishing magnetic field we assume an equidistant
mean-field spectrum with unit spacing.  The two-body random interactions have
fixed couplings $C_0 = C_1=1/10$.  We add the Zeeman Hamiltonian (\ref{Z}) to
the system and increase the Zeeman splitting $2\mu B$ from zero to its maximal
value $\langle\varepsilon_{i+1}-\varepsilon_i\rangle/2$.  Simultaneously, the
variance of the imaginary part of the random matrix elements is increased from
zero to one, being held proportional to the Zeeman splitting. Figure~\ref{fig4}
shows that the odd-even staggering decreases with increasing Zeeman splitting.
The remaining staggering is due to the interactions, which are relatively weak
in this example; the transition from the GOE to the GUE in the random two-body
matrix is very mild.  For strong two-body interactions the odd-even staggering
remains strong when time-reversal symmetry is broken.  Thus, the breaking of
time-reversal invariance has only mild effects on the ground-state structure in
strongly interacting systems.  This finding is consistent with a recent study
of time-reversal symmetry breaking in the nuclear shell model with random
two-body interactions \cite{BFP}. 

\section{Summary}
We have shown analytically and numerically that random two-body interactions
cause an odd-even staggering in interacting few-fermion systems such as small
metallic grains or quantum dots.  Interactions tend to smooth out the odd-$N$
and even-$N$ dependence of the pairing gaps and can thereby be discriminated
from the non-smooth mean-field staggering.  As expected, the breaking of time
reversal symmetry leads to a decrease of the odd-even staggering; this trend
can however be countered by sufficiently strong two-body interactions.

\section*{Acknowledgments}
This research used resources of the Center for Computational Sciences
at Oak Ridge National Laboratory (ORNL). T.~P. acknowledges support as
a Wigner Fellow and staff member at ORNL, managed by UT-Battelle, LLC
for the U.S. Department of Energy under Contract DE-AC05-00OR22725.
L.~K.  and G.~F.~B. are likewise supported by the U.S. Department of
Energy, under Grant DE-FG03-00ER41132.

\appendix \section*{}

The derivation of interaction-induced staggering in Section~\ref{2body} was
obtained in the dilute limit $N \ll M$. For general values of $N$ and $M$ and
couplings $C_0$ and $C_1$, a straightforward counting procedure results in
the exact expressions
\begin{eqnarray}
{\rm Tr}&_{J=0}& H^2 =
{C_0^2+3C_1^2 \over 64} N^2(2M-N)^2 \nonumber \\ &+& {N \over 16}
[C_0^2(2M^2+MN-N^2) \nonumber \\& &\;\;\;-3C_1^2(2M^2-7MN+3N^2)] \nonumber \\
&+& {N \over 16} [C_0^2(6M+N-2Nd(1-d)) \nonumber
\\ & & \;\;\;-3C_1^2(10M-13N-2Nd(1+d))]
\nonumber \\
&+& {N \over 16} \left[8C_0^2-24C_1^2(2+d)\right]  \label{tr0} \\
{\rm Tr}&_{J=1/2}& H^2 =
{C_0^2+3C_1^2 \over 64} N^2(2M-N)^2 \nonumber \\ &+& {N \over 16}
[C_0^2(2M^2+MN-N^2) \nonumber \\& &\;\;\;-3C_1^2(2M^2-7MN+3N^2)] \nonumber \\
&+& {1 \over 16} [C_0^2(-3M^2+9MN-N^2/2) \nonumber
\\ & & \;\;\;+3C_1^2(M^2-15MN+31N^2/2+2N^2d(1+d))] \nonumber \\
&+& {1 \over 16}[C_0^2(-9M+11N-2Nd) \nonumber \\&+&3C_1^2(9M-31N-8Nd)]
\nonumber \\ &+&{3 \over 64}[-13C_0^2+73C_1^2] \label{tr1} \\
{\rm Tr}&_{J=N/2}& H^2 =
{C_1^2 \over 4}[N^2(M-N)^2-N(M^2-5MN+4N^2) \nonumber \\
&+& N(7N-3M)-4N] \label{trn}
\end{eqnarray}

In each of the three above expressions for minimal and maximal spin states,
terms are ordered by their relative importance in the many-body limit at finite
density, i.e., in the limit $N \to \infty$ with $\rho \equiv N/2M = const$.
The leading term is $O(N^4)$ in the many-body limit, and this leading term is
seen to be manifestly symmetric under particle-hole exchange $N \to 2M -N$ for
the minimal-spin states (of course the maximal spin states $J=N/2$ exist only
for $N\le M$). At subleading order, the symmetry is broken due to
anticommutation relations between the creation and annihilation operators in
Eq.~(\ref{hamilt}). At both leading and first subleading order, ${\rm Tr} H^2$
is clearly identical for the $J=0$ and $J=1/2$ states, indicating that the
staggering can occur only at $O(1/N^2)$, entirely consistent with our dilute
analysis in Section~\ref{2body}.  It is also at this second subleading order
that we first encounter the dimensionless quantity $d$, which we did not need
to consider in the dilute approximation.  $d$, taking values $0 \le d \le 1$,
represents the fraction of particles in the basis state that live on doubly
occupied orbitals.

As discussed above, for a pure $S=0$ coupling ($C_1$ vanishing), ground states
come always from the sector of minimal spin, and thus we are led to 
consider the quantity
\begin{eqnarray}
{{\rm Tr}_{S=0} H^2 -{\rm Tr}_{S=1/2} H^2 \over {\rm Tr}_{S=1/2} H^2 }
&=& {3-6\rho(1-\rho)-8\rho^2 d(1-d) \over N^2(1-\rho)^2} \nonumber \\
& \ge& {3 \over 2 N^2}\,,
\label{c0diff}
\end{eqnarray}
where terms of higher order in the $1/N$ expansion have been dropped, and the
last inequality is easily checked for all possible values of filling fraction
$\rho$ and double occupancy fraction $d$. Thus, our original estimate,
Eq.~(\ref{dilutedelta}), obtained using the dilute approximation, is confirmed
as a lower bound to the amount of predicted pairing gap,
\begin{equation}
(-1)^N \Delta(N)_{C_0} \ge {3 \over 2 N^2}|E(N)| \,.
\end{equation}

The situation is more complex for a pure $S=1$ coupling ($C_0=0$), since here
the ground state may be a state of either minimal or maximal spin, depending on
the density $\rho$. Comparing Eqs.~(\ref{tr0},\ref{tr1}) with Eq.~(\ref{trn})
at leading order in the many-body limit, we see easily that $J=N/2$ is
preferred at very low density, $\rho < \rho_{crit}= (5-2\sqrt{3})/13 \approx
0.118$, but as density increases a transition should occur to ground states of
minimal spin. The preference for maximal spin at low density is obvious, since
high-spin states clearly maximize the fraction of particle pairs with aligned
spins ($S=1$ instead of $S=0$). On the other hand, the physical reason for the
transition to minimal-spin ground states even with a pure $S=1$ coupling for
$\rho>\rho_{crit}$ is that at high enough density there are relatively few
other high-spin states that a given high-spin state can couple to.

The relevant result for our purpose here is that at low densities there is no
predicted stagger in the many-body limit, in accordance with Eq.~(\ref{trn}),
but for $\rho>\rho_{crit}$ minimal-spin states again become dominant. A
calculation completely analogous to the one in Eq.~(\ref{c0diff}) tells us that
once again the pairing gap $\Delta$ is positive (negative) for even (odd) $N$
and proportional to $1/N^2$ times the magnitude of the binding energy. Thus,
\begin{eqnarray}
\Delta(N)_{C_1} &=&0
\;\;\;\;\;\;\;\;\;\;\;\;\;\;\;\;\;\;\;\;\;\;\;
 (\rho< \rho_{crit}) \nonumber \\
(-1)^N\Delta(N)_{C_1} &\ge& {0.027 \over N^2}|E(N)| \;\;\;\;\; (\rho>
\rho_{crit}) \end{eqnarray}

The above analysis generalizes easily to the generic case where the two
coefficients $C_0$ and $C_1$ are both nonzero. For $C_0>C_1$, ground states are
always expected to come from the minimal-spin sector, leading to positive
pairing gap $\Delta$ proportional to $1/N^2$ times the binding energy.  For
$C_1>C_0$, on the other hand, there will be a transition between no pairing gap
at low density to positive pairing gap at higher density, the critical density
$\rho_{crit}$ approaching $0.118$ for $C_1 \gg C_0$ and approaching $0$ at $C_0
=C_1$.

\end{multicols}

\end{document}